# Experimental optimal verification of three-dimensional entanglement on a silicon chip

Lijun Xia[1,5], Liangliang Lu[2,1,5], Kun Wang[3,1], Xinhe Jiang[4,1] , Shining Zhu[1] and Xiaosong Ma[1,6]

[1] National Laboratory of Solid-state Microstructures, School of Physics, Collaborative Innovation Center of Advanced Microstructures, State Key Laboratory for Novel Software Technology, Department of Computer Science and Technology, Nanjing University, Nanjing 210093, China.
[2] Key Laboratory of Optoelectronic Technology of Jiangsu Province,
School of Physical Science and Technology, Nanjing Normal University, Nanjing 210023, China.
[3] Institute for Quantum Computing, Baidu Research, Beijing 100193, China.
[4] Institute for Quantum Optics and Quantum Information, Austrian Academy of Sciences, 1090 Vienna, Austria.
[5] These authors contributed equally: Lijun Xia, Liangliang Lu.
[6]Author to whom any correspondence should be addressed.

E-mail: Xiaosong.Ma@nju.edu.cn

**Abstract**

High-dimensional entanglement is significant for the fundamental studies of quantum physics and offers unique advantages in various quantum information processing (QIP) tasks. Integrated quantum devices have recently emerged as a promising platform for creating, processing, and detecting complex high-dimensional entangled states. A crucial step towards practical quantum technologies is to verify that these devices work reliably with an optimal strategy. In this work, we experimentally implement an optimal quantum verification strategy on a three-dimensional maximally entangled state using local projective measurements on a silicon photonic chip. A 95% confidence is achieved from 1190 copies to verify the target quantum state. The obtained scaling of infidelity as a function of the number of copies is -0.5497±0.0002, exceeding the standard quantum limit of -0.5 with 248 standard deviations. Our results indicate that quantum state verification could serve as an efficient tool for complex quantum measurement tasks.

Keywords: quantum optics, silicon chip, high-dimensional entanglement, quantum state verification

## 1. Introduction

An entangled state is an essential resource in a variety of QIP tasks. High-dimensional entangled states are of especial interest, owing to their distinctive properties compared to qubit states. They enable larger channel capacity and better noise tolerance in quantum communication [1-3], higher resolution of quantum sensing[4], greater efficiency and flexibility in quantum computing [5, 6], and a richer variety of quantum simulations [7, 8].

Experimental realizations of high-dimensional entanglement with photons have demonstrated with various degrees of freedom on different platforms [9-20]. Among these works, path entangled photon pairs generated on integrated quantum photonic chips are particularly attractive, due to the conceptual simplicity and the excellent scalability. Preferably, path encoded high-dimensional states are often configured by silicon-based photonic integrated circuits, due to their small footprint, reduced power consumption, and enhanced processing stability, which is challenging in bulk optical designs [21-24]. Moreover, silicon photonic devices are compatible with complementary metal-oxide-semiconductor (COMS) fabrication.

An efficient method to verify the high-dimensional entangled state is highly desirable. Traditionally, reconstruction of a state is realized by quantum state tomography (QST) [25]; however, it is time- and computation-consuming and also challenging for statistical analysis. Non-tomographic approaches have been proposed by incorporating some assumptions either on the quantum states or on the available operations [26-33]. In many application scenarios, it is only required to verify that the produced states are sufficiently close to the target state. Optimal quantum state verification (QSV) [34] is a procedure that aims at devising efficient protocols for verifying the target states through measurements that can be realized by local operations and classical communication [35]. Up to now, efficient QSV protocols for bipartite [36-39] or multipartite pure states [40, 41] have been proposed and implemented using both nonadaptive [42] and adaptive measurements [43, 44]. The optimal verification protocol from two-qubit entangled states has been extended to maximally entangled states for any finite dimension [45] using locally projective, two-outcome measurements. Note that entanglement certification [46-48] and efficient entanglement verification has been proposed [49] and experimentally realized [50]. Very recently, efficient experimental verification of quantum gates with local operations was also realized [51].

In this work, a silicon photonic chip is employed both to generate and to verify a maximally path-entangled qutrits state by using an optimal nonadaptive strategy. Our results indicate that QSV can offer a precise estimation of the reliability of quantum devices and serve as a standardized procedure to efficiently verify complex integrated entangled quantum states.

## 2. Verification procedure and experimental setup

Here we introduce the quantum state verification protocol. Suppose a quantum device is expected to produce a target state $|\Psi\rangle$, whereas it produces $\sigma_1, \sigma_2, \ldots \sigma_N$ independently in N runs. The task of QSV is to distinguish either $\sigma_i = |\Psi\rangle$ for all i, or $\langle\Psi|\sigma_i|\Psi\rangle > 1 - \varepsilon$ for all i, i.e., the produced states are far from the target state $|\Psi\rangle$ by ε. Ideally, the optimal strategy is to project $\sigma_i$ to the space of target state $|\Psi\rangle$ and its orthogonal space. However, this would require entangled measurements in general, especially when the target state is entangled. Protocols based on local projective measurements are more feasible and experiment-friendly in practice. In this work, binary-outcome measurements perform from a set of accessible measurements. Each two-outcome measurement $\{M_j, 1 - M_j\}$ (j = 1, 2, 3, ...) is specified by an operator with probability $p_j$, satisfying $M_j|\Psi\rangle = |\Psi\rangle$, corresponding to pass the test. The maximal probability that $\sigma_i$ can pass the test is given by [34]

$$\max_{\langle\Psi|\sigma_i|\Psi\rangle \leq 1-\varepsilon} Tr(\Omega\sigma_i) = 1 - [1 - \lambda_2(\Omega)]\varepsilon = 1 - \Delta_\varepsilon, \tag{1}$$

where $\Omega = \Sigma_j p_j M_j$ is a verification strategy, $\lambda_2(\Omega)$ is the second largest eigenvalue of $\Omega$ and $\Delta_\varepsilon$ is the rejection probability of a single test. After N runs, the incorrect state $\sigma_i$ can pass the test with probability being at most $[1 - [1 - \lambda_2(\Omega)]\varepsilon]^N$. To guarantee confidence $1 - \delta$, the minimum number of N must be

$$N = \frac{1}{[1-\lambda_2(\Omega)]\varepsilon} ln\frac{1}{\delta}. \tag{2}$$

Eq. (2) implies that the optimal strategy within the set of accessible measurements is to minimize $\lambda_2(\Omega)$. Recently, Li et al. [38] and Zhu et al. [45] proposed the optimal protocols for maximally entangled states using complete sets of mutually unbiased bases (MUB). For the qudit case, a complete set of MUBs has d + 1 elements. Each element can be written as $\{|\varphi_{0,i}\rangle, |\varphi_{1,i}\rangle, \ldots |\varphi_{d-1,i}\rangle\}$ depending on measurement $S_i$. For a qutrit case, d = 3 and the MUBs can be defined as

$$S_1 : \{(1,0,0), (0,1,0), (0,0,1)\}, \tag{3}$$

$$S_2 : \left\{\frac{1}{\sqrt{3}}(1,1,1), \frac{1}{\sqrt{3}}(1,\omega,\omega^*), \frac{1}{\sqrt{3}}(1,\omega^*,\omega)\right\}, \tag{4}$$

$$S_3 : \left\{\frac{1}{\sqrt{3}}(1,\omega,1), \frac{1}{\sqrt{3}}(1,\omega^*,\omega^*), \frac{1}{\sqrt{3}}(1,1,\omega)\right\}, \tag{5}$$

$$S_4 : \left\{\frac{1}{\sqrt{3}}(1,\omega^*,1), \frac{1}{\sqrt{3}}(1,1,\omega^*), \frac{1}{\sqrt{3}}(1,\omega,\omega)\right\}, \tag{6}$$

where $\omega = e^{i2/3\pi}$. To achieve confidence $1-\delta$, the minimum number of measurements required is $\frac{1}{\Delta_\varepsilon} ln\frac{1}{\delta}$, with $\Delta_\varepsilon = \frac{3}{4}\varepsilon$. The optimal MUB strategy for verifying bipartite maximally entangled state $|\Psi\rangle = \frac{1}{\sqrt{d}}\sum_{j=0}^{d-1}|jj\rangle$ is as follows: Alice randomly chooses a measurement $S_i$ with probability $1/(d+1)$, while Bob performs the projective measurement in the conjugate basis $S_i^*$. The test is passed if the state of Alice and Bob is projected on $|\varphi_{k,i}\rangle|\varphi_{k,i}^*\rangle$ for any k, where $|\varphi_{k,i}^*\rangle$ denotes the complex conjugate of $|\varphi_{k,i}\rangle$ with respect to the standard bases [38, 45, 52].

Practically, the produced states often have a limited fidelity to the target state, which can result in an occasional rejection within a few amounts measurements, incurring an incorrect conclusion of QSV. As shown in Fig. 1, we consider a more practical protocol called quantum fidelity estimation based on a sufficient amount of copies (N) and use the relative frequency of passing copies (m) to describe the device in a statistical way [37, 42, 43, 52]. For the bipartite states $\sigma_i$ generated by the practical quantum device, the task is to distinguish the following two cases (also called hypotheses):

Case 1: the *average* fidelity of the quantum device's produced states is larger than $1-\varepsilon$: $\frac{1}{N}\sum_{i=1}^{N}\langle\Psi|\sigma_i|\Psi\rangle > 1-\varepsilon$.

Case 2: the *average* fidelity of the quantum device's produced states is less than $1-\varepsilon$: $\frac{1}{N}\sum_{i=1}^{N}\langle\Psi|\sigma_i|\Psi\rangle \leq 1-\varepsilon$.

As shown in Fig.1, a conclusion can be drawn as the states are within ($\sigma_i \in \bar{B}$, Case 1) the ε-target circle on average with a certain confidence. Intuitively, when the number of passing tests m is larger than $N(1-\Delta_\epsilon)$, the device belongs to Case 1 with high probability, otherwise it would be very unlikely to obtain the experimental data. Quantitatively, the confidence $1-\delta$ that the device belongs to Case 1 can be determined using the Chernoff bound [35, 42-44, 49, 50, 52]:

$$e^{-ND(\frac{m}{N}\|1-\Delta_\varepsilon)} \equiv \delta, \tag{7}$$

where

$$D(x\|y) \equiv x\, log_2\frac{x}{y} + (1-x)\, log_2\frac{1-x}{1-y}, \tag{8}$$

is Kullback-Leibler divergence. Notice that the value $\delta$ measures how unlikely the given data are if Case 2 is true. We generate many copies of $\sigma_i$, and apply the optimal conjugate strategy sequentially according to the corresponding projectors. From the measured data, a single coincidence count can be obtained for each randomly chosen measurement setting and decide whether $\sigma_i$ passes the test or not. We increase the number of copies and obtain the passing probability m/N. The values of $\delta(\varepsilon)$ are obtained from Eq. (7) given certain $\varepsilon(\delta)$.

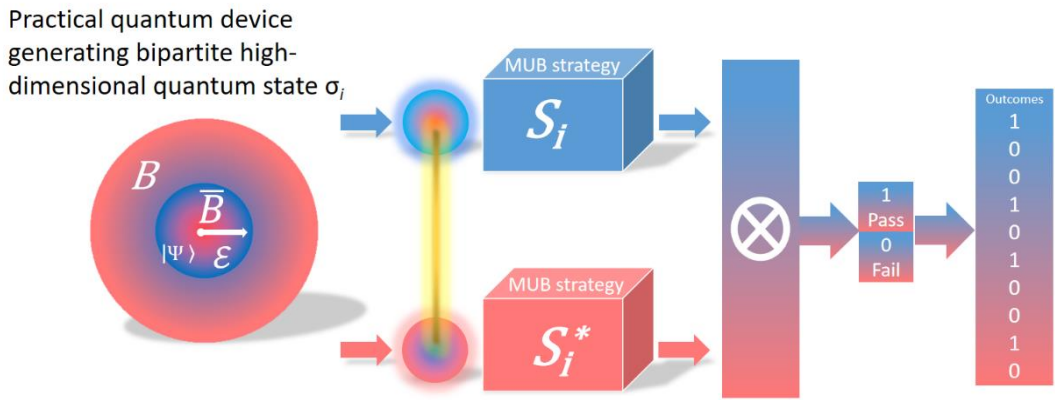

**Fig. 1** Optimal verification of maximally entangled three-dimensional state $|\Psi\rangle = 1/\sqrt{d}\sum_{j=0}^{d-1}|jj\rangle$. Alice randomly chooses a projective measurement $S_i$ (with i = 1, … , d + 1) from the complete set of MUB with probability d + 1, while Bob performs the projective measurement in the conjugate basis. Each measurement returns a binary outcome 1 or 0, associated with pass or fail of the test, respectively. After N runs, the protocol returns m passing outcomes, giving a passing probability of m/N.

Here we experimentally realize the generation and verification of a pair of entangled qutrits on a silicon chip (Fig.2a) encompassing three dual Mach-Zehnder-interferometer rings (DMZI-Rs) [53-55]. Signal (1549.5 nm) and idler (1558.3 nm) photons are generated by annihilating two pump photons (1552.1 nm) in non-degenerate spontaneous four wave mixing (SFWM) in the DMZI-R sources, and further separated by on-chip wavelength division multiplexers (WDMs). This produces a three-dimensional state of the form $\Psi = \sum_{k=0}^{2} C_k |k\rangle|k\rangle$, the coefficient $C_k$ can be arbitrarily changed by adjusting pump power and phase over the sources. With a balanced generation rate for all three sources and zero relative phases of the pump, we generate a maximally entangled state of two qutrits $|\Psi\rangle = \frac{1}{\sqrt{3}}(|00\rangle + |11\rangle + |22\rangle)$, where $|0\rangle$, $|1\rangle$, and $|2\rangle$ represents the individual path states of single photons. Each qutrit can be locally manipulated by reconfigurable linear optical circuits for implementing arbitrary 3-D unitary operations via three-dimensional multiports (3D-MPs). 3D-MPs are composed of 28 phase shifters and 22 multimode interferometers (MMIs) to realize $R_z(\varphi_z)$ and $R_y(\theta_z)$ rotations, respectively.

The full scheme of our experiment is shown in Fig.2b. A tunable picosecond fiber laser produces pulses with 7.8 ps duration, 60.2 MHz repetition rate, and -0.8dBm average power. A bandpass filter with 1.2 nm bandwidth then suppresses the unwanted amplified spontaneous emission (ASE) noise of the laser for ~ 40 dB. Before the light is coupled via transverse electric (TE) grating couplers into the chip, a polarization controller (PC) optimize the polarization of the pump for maximizing fiber-to-chip. Off-chip single-channel WDMs filter entangled photons emerging from the chip to remove the residual pump. Six superconducting single photon detectors (SSPDs) detect photons filtered after WDMs with ~80% detection efficiency, ~100 Hz dark count rates, a ~ 50 ns dead time. A field-programmable gate array (FPGA)-based time tag device collects electrical signals of the detector. The losses for the photons in the path entanglement measurement add up to 18.73 dB~19.13 dB. A programmable current source controls all the heaters with a range of 0-20 mA and 16-bit resolution. The coincidence counts $CC_{ij}$ between path i (i = 1, 3, 5) and path j (j = 2, 4, 6) are extracted as the experimental results. Finally, we verify the qutrit entanglement by analyzing coincidence counts from the outputs.

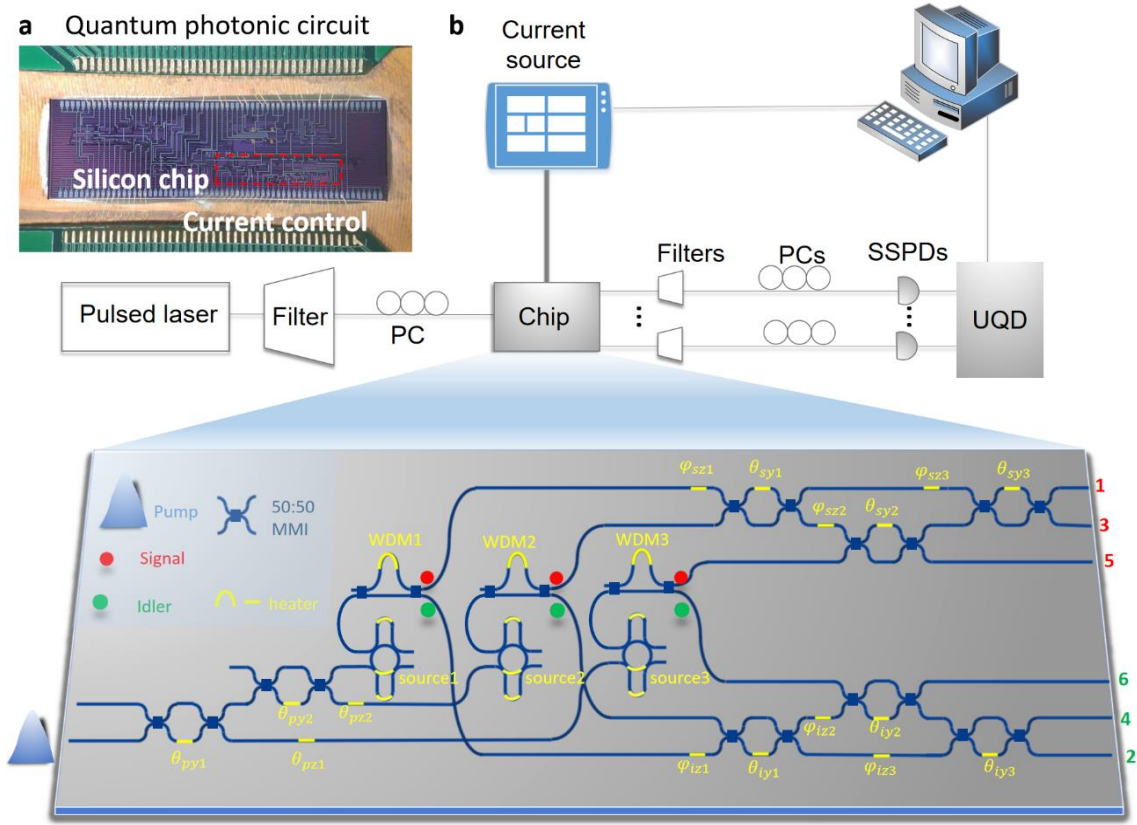

**Fig.2 a** Optical microscope image of our quantum photonic circuit. **b** Schematic of the complete experimental setup. A picosecond pump pulse is filtered, polarized and coupled into the chip by a grating coupler. A photon pair is created in a superposition among three coherently pumped sources, generating a maximally entangled qutrit state. The signal (red) and idler (blue) photons are separated by unbalanced Mach-Zehnder interferometer (MZI) and routed to two universal multiport interferometers, enabling arbitrary local unitary transformation. Photons coupled out of the chip are polarization optimized and separated from residual pump by filters before detected by six SSPDs. Coincidence events are recorded by the time tag device. Phase shifters on the device are configured through current sources controlled by a computer.

## 3. Results

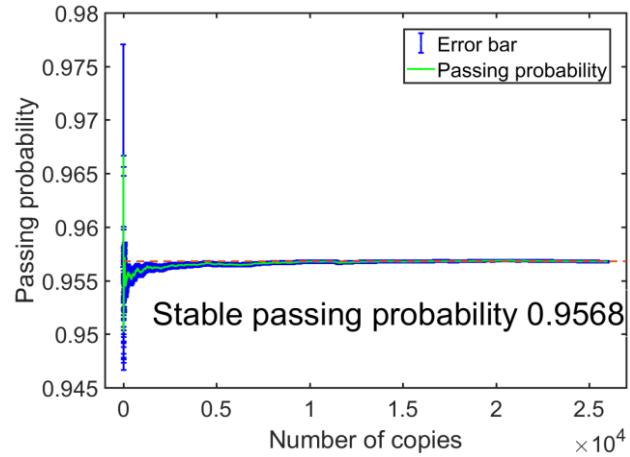

**Fig. 3** Experimental results on the variation of passing probability ( m/N ) versus the number of copies. The passing probability will reach a stable value 0.9568. The blue symbol is the experimental error bar, which is obtained by 300 trials of measurement.

We give the variation of extracted passing probability versus the number of copies in Fig. 3. The passing probability reaches a stable value 0.9568. The fidelity between $\sigma_i$ and $|\Psi\rangle\langle\Psi|$, which can be inferred from the asymptotic passing probability via Eq. (1), is about 94.24% assuming that the number of copies is large enough so that m/N approximates $1 - \Delta_\varepsilon$. This result indicates that the quantum device is stable. In Fig. 4 we present the results of applying QSV to estimate the average fidelity of the quantum states generated by the quantum device. In Fig. 4, δ is obtained by setting infidelity ε as 0.08 (Case 1). Fig. 4 shows that within N=1190 copies, the passing probability $m/N = 0.9563$ and δ approaches 0.05. We thus conclude that the generated states have an average fidelity larger than 0.92 with 95% confidence. The values of δ are obtained from Eq. (7) and plotted in log-scale in the insets of Fig. 4. For the estimation of confidence in verifying our target state in Case 1, these few amounts of copies required in our protocol are useful when only limited resources are available in practical quantum device.

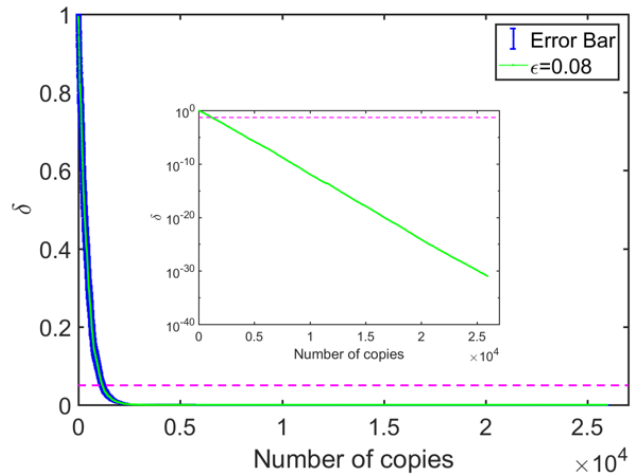

**Fig. 4** Experimental results for the QSV. When infidelity ε is set to be 0.08, δ is plotted as a function of the number of copies, where 1-δ is the confidence of the device belonging to Case 1. δ approaches 0.05 within 1190 copies for Case 1. Insets show δ versus the number of copies in log-scale.

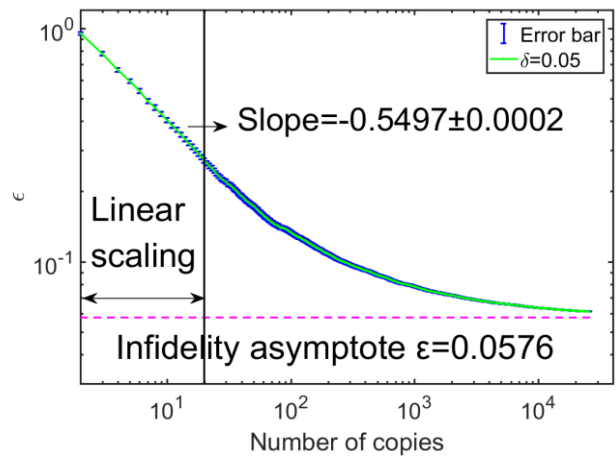

**Fig. 5** The variation of infidelity parameter vs. the number of copies. The parameter ε is log-log plotted versus the number of copies with δ = 0.05. The blue symbol is the experimental error bar, which is obtained by 300 trials of measurements. The error of the slope is obtained by fitting 100 groups of ε-N data.

In Fig. 5, we set $\delta = 0.05$ and calculate the infidelity ε as a function of the number of copies. In the log-log plot, ε drops fast at small number of copies and converges to 0.0576. The slope obtained from fitting the linear part is approximately -0.5497±0.0002. The error of the slope is obtained by fitting 100 groups of ε-N data. The non-optimal scaling of ε versus $1/N$ is due to the limited fidelity of the generated state. This scaling obviously exceeds the standard quantum bound of -0.5 by 248 standard deviations [42, 43], which is obtained by subtracting the bound (-0.5) from the fitting value (-0.5497) and dividing it by the error (0.0002). This result indicates that QSV is an efficient quantum measurement tool for high-dimensional entangled states. It is worthy to note that the slope can approach the Heisenberg limit of $-1$ in quantum metrology if the fidelity is further improved [43].

## 4. Conclusion

In summary, we experimentally demonstrate the optimal verification of maximally entangled qutrit state generated on a silicon chip with local projective measurements. The variation of confidence and infidelity parameters with the number of copies are presented. We give a comprehensive analysis of the generated state and present a precise estimation of the reliability and stability of the device. The results provide a scaling parameter better than the value of the standard quantum limit. QSV represents a compelling technique for quantifying the prepared state with substantially lower complexity than the quantum state tomography. Measurement settings of chip-based verification protocol can be implemented with programmable and reconfigurable devices. The present procedure for chip-based states verification can be extended to higher-dimensional and multipartite on-chip entangled states, requiring good scalability in the integrated platform. Our work paves the way for the future realization of large-scale chip-based high-dimensional quantum states verification.

**Acknowledgements**

This research is supported by the National Key Research and Development Program of China (2017YFA0303704, 2019YFA0308700), National Natural Science Foundation of China (Grants No. 11690032, No.9 11321063 and 12033002), NSFC-BRICS (No. 61961146001), Leading-edge technology Program of Jiangsu Natural Science Foundation (BK20192001), and the Fundamental Research Funds for the Central Universities.

**Data availability statement**

The data that support the findings of this study are available upon reasonable request from authors.

**References**

[1]　Gisin N, Ribordy G, Tittel W and Zbinden H 2002 Quantum cryptography *Rev. Mod. Phys.* **74** 145-95. https://doi.org/10.1103/RevModPhys.74.145
[2]　D'Ambrosio V, Nagali E, Walborn S P, Aolita L, Slussarenko S, Marrucci L and Sciarrino F 2012 Complete experimental toolbox for alignment-free quantum communication *Nat. Commun.* **3** 961. https://doi.org/10.1038/ncomms1951
[3]　Graham T M, Bernstein H J, Wei T-C, Junge M and Kwiat P G 2015 Superdense teleportation using hyperentangled photons *Nat. Commun.* **6** 7185. https://doi.org/10.1038/ncomms8185
[4]　Fickler R, Lapkiewicz R, Plick W N, Krenn M, Schaeff C, Ramelow S and Zeilinger A 2012 Quantum entanglement of high angular momenta *Science* **338** 640-3. https://doi.org/10.1126/science.1227193
[5]　Lanyon B P*, et al.* 2009 Simplifying quantum logic using higher-dimensional hilbert spaces *Nat. Phys.* **5** 134-40. https://doi.org/10.1038/NPHYS1150
[6]　Qiang X G*, et al.* 2018 Large-scale silicon quantum photonics implementing arbitrary two-qubit processing *Nat. Photon.* **12** 534-9. https://doi.org/10.1038/s41566-018-0236-y
[7]　Karnieli A and Arie A 2018 Frequency domain stern-gerlach effect for photonic qubits and qutrits *Optica* **5** 1297-303. https://doi.org/10.1364/OPTICA.5.001297
[8]　Lu H H*, et al.* 2019 Simulations of subatomic many-body physics on a quantum frequency processor *Phys. Rev. A* **100** 012320. https://doi.org/10.1103/PhysRevA.100.012320
[9]　Mair A, Vaziri A, Weihs G and Zeilinger A 2001 Entanglement of the orbital angular momentum states of photons *Nature* **412** 313-6. https://doi.org/10.1038/35085529
[10]　Wang X L, Cai X D, Su Z E, Chen M C, Wu D, Li L, Liu N L, Lu C Y and Pan J W 2015 Quantum teleportation of multiple degrees of freedom of a single photon *Nature* **518** 516-9. https://doi.org/10.1038/nature14246
[11]　Liu S L*, et al.* 2018 Coherent manipulation of a three-dimensional maximally entangled state *Phys. Rev. A* **98** 062316. https://doi.org/10.1103/PhysRevA.98.062316
[12]　Kues M*, et al.* 2017 On-chip generation of high-dimensional entangled quantum states and their coherent control *Nature* **546** 622. https://doi.org/10.1038/nature22986
[13]　Imany P, Jaramillo-Villegas J A, Odele O D, Han K H, Leaird D E, Lukens J M, Lougovski P, Qi M H and Weiner A M 2018 50-ghz-spaced comb of high-dimensional frequency-bin entangled photons from an on-chip silicon nitride microresonator *Opt. Express* **26** 1825-40. https://doi.org/10.1364/OE.26.001825
[14]　Guo Y, Hu X M, Liu B H, Huang Y F, Li C F and Guo G C 2018 Experimental realization of path-polarization hybrid high-dimensional pure state *Opt. Express* **26** 28918-26. https://doi.org/10.1364/OE.26.028918
[15]　Wang J W*, et al.* 2018 Multidimensional quantum entanglement with large-scale integrated optics *Science* **360** 285-91. https://doi.org/10.1126/science.aar7053
[16]　Schaeff C, Polster R, Huber M, Ramelow S and Zeilinger A 2015 Experimental access to higher-dimensional entangled quantum systems using integrated optics *Optica* **2** 523-9. https://doi.org/10.1364/OPTICA.2.000523
[17]　Jin X M, Peng C Z, Deng Y J, Barbieri M, Nunn J and Walmsley I A 2013 Sequential path entanglement for quantum metrology *Sci. Rep.* **3** 1779. https://doi.org/10.1038/srep01779
[18]　Solntsev A S and Sukhorukov A A 2017 Path-entangled photon sources on nonlinear chips *Reviews in Physics* **2** 19-31. https://doi.org/https://doi.org/10.1016/j.revip.2016.11.003
[19]　Marcikic I, de Riedmatten H, Tittel W, Scarani V, Zbinden H and Gisin N 2002 Time-bin entangled qubits for quantum communication created by femtosecond pulses *Phys. Rev. A* **66** 062308. https://doi.org/10.1103/PhysRevA.66.062308
[20]　Tiranov A, Designolle S, Cruzeiro E Z, Lavoie J, Brunner N, Afzelius M, Huber M and Gisin N 2017 Quantification of multidimensional entanglement stored in a crystal *Phys. Rev. A* **96** 040303. https://doi.org/10.1103/PhysRevA.96.040303
[21]　Reimer C*, et al.* 2019 High-dimensional one-way quantum processing implemented on d-level cluster states *Nat. Phys.* **15** 148-53. https://doi.org/10.1038/s41567-018-0347-x
[22]　Reck M, Zeilinger A, Bernstein H J and Bertani P 1994 Experimental realization of any discrete unitary operator *Phys. Rev. Lett.* **73** 58-61. https://doi.org/10.1103/PhysRevLett.73.58
[23]　Schaeff C, Polster R, Lapkiewicz R, Fickler R, Ramelow S and Zeilinger A 2012 Scalable fiber integrated source for higher-dimensional path-entangled photonic qunits *Opt. Express* **20** 16145-53. https://doi.org/10.1364/OE.20.016145
[24]　Silverstone J W*, et al.* 2014 On-chip quantum interference between silicon photon-pair sources *Nat. Photon.* **8** 104-8. https://doi.org/10.1038/NPHOTON.2013.339
[25]　Kent A 1999 Noncontextual hidden variables and physical measurements *Phys. Rev. Lett.* **83** 3755-7. https://doi.org/10.1103/PhysRevLett.83.3755
[26]　Toth G and Guhne O 2005 Detecting genuine multipartite entanglement with two local measurements *Phys. Rev. Lett.* **94** 060501. https://doi.org/10.1103/PhysRevLett.94.060501
[27]　Flammia S T and Liu Y K 2011 Direct fidelity estimation from few pauli measurements *Phys. Rev. Lett.* **106** 230501. https://doi.org/10.1103/PhysRevLett.106.230501

[28] da Silva M P, Landon-Cardinal O and Poulin D 2011 Practical characterization of quantum devices without tomography *Phys. Rev. Lett.* **107** 210404. https://doi.org/10.1103/PhysRevLett.107.210404
[29] Aolita L, Gogolin C, Kliesch M and Eisert J 2015 Reliable quantum certification of photonic state preparations *Nat. Commun.* **6** 8498. https://doi.org/10.1038/ncomms9498
[30] Hayashi M and Morimae T 2015 Verifiable measurement-only blind quantum computing with stabilizer testing *Phys. Rev. Lett.* **115** 220502. https://doi.org/10.1103/PhysRevLett.115.220502
[31] McCutcheon W, *et al.* 2016 Experimental verification of multipartite entanglement in quantum networks *Nat. Commun.* **7** 13251. https://doi.org/10.1038/ncomms13251
[32] Takeuchi Y and Morimae T 2018 Verification of many-qubit states *Phys. Rev. X* **8** 021060. https://doi.org/10.1103/PhysRevX.8.021060
[33] Badescu C, O'Donnell R and Wright J 2019 Quantum state certification. In: *PROCEEDINGS OF THE 51ST ANNUAL ACM SIGACT SYMPOSIUM ON THEORY OF COMPUTING (STOC '19),* pp 503-14
[34] Pallister S, Linden N and Montanaro A 2018 Optimal verification of entangled states with local measurements *Phys. Rev. Lett.* **120** 170502. https://doi.org/10.1103/PhysRevLett.120.170502
[35] Morris J, Saggio V, Gočanin A and Dakić B 2021 Quantum verification with few copies *arXiv e-prints* arXiv:2109.03860. https://doi.org/arXiv:2109.03860
[36] Wang K and Hayashi M 2019 Optimal verification of two-qubit pure states *Phys. Rev. A* **100** 032315. https://doi.org/10.1103/PhysRevA.100.032315
[37] Yu X D, Shang J W and Guhne O 2019 Optimal verification of general bipartite pure states *npj Quantum Inf* **5** 112. https://doi.org/10.1038/s41534-019-0226-z
[38] Li Z H, Han Y G and Zhu H J 2019 Efficient verification of bipartite pure states *Phys. Rev. A* **100** 032316. https://doi.org/10.1103/PhysRevA.100.032316
[39] Zhu H J and Hayashi M 2019 Efficient verification of pure quantum states in the adversarial scenario *Phys. Rev. Lett.* **123**https://doi.org/10.1103/PhysRevLett.123.260504
[40] Liu Y C, Yu X D, Shang J W, Zhu H J and Zhang X D 2019 Efficient verification of dicke states *Phys. Rev. Applied* **12** 044020. https://doi.org/10.1103/PhysRevApplied.12.044020
[41] Li Z H, Han Y G and Zhu H J 2020 Optimal verification of greenberger-horne-zeilinger states *Phys. Rev. Applied* **13** 054002. https://doi.org/10.1103/PhysRevApplied.13.054002
[42] Zhang W H, *et al.* 2020 Experimental optimal verification of entangled states using local measurements *Phys. Rev. Lett.* **125** 030506. https://doi.org/10.1103/PhysRevLett.125.030506
[43] Jiang X H, *et al.* 2020 Towards the standardization of quantum state verification using optimal strategies *npj Quantum Inf* **6** 90. https://doi.org/10.1038/s41534-020-00317-7
[44] Zhang W H, *et al.* 2020 Classical communication enhanced quantum state verification *npj Quantum Inf* **6** 103. https://doi.org/10.1038/s41534-020-00328-4
[45] Zhu H J and Hayashi M 2019 Optimal verification and fidelity estimation of maximally entangled states *Phys. Rev. A* **99** 052346. https://doi.org/10.1103/PhysRevA.99.052346
[46] Knips L, Schwemmer C, Klein N, Wiesniak M and Weinfurter H 2016 Multipartite entanglement detection with minimal effort *Phys. Rev. Lett.* **117** 210504. https://doi.org/10.1103/PhysRevLett.117.210504
[47] Bavaresco J, Valencia N H, Klockl C, Pivoluska M, Erker P, Friis N, Malik M and Huber M 2018 Measurements in two bases are sufficient for certifying high-dimensional entanglement *Nat. Phys.* **14** 1032-7. https://doi.org/10.1038/s41567-018-0203-z
[48] Friis N, Vitagliano G, Malik M and Huber M 2019 Entanglement certification from theory to experiment *Nat. Rev. Phys.* **1** 72-87. https://doi.org/10.1038/s42254-018-0003-5
[49] Dimic A and Dakic B 2018 Single-copy entanglement detection *npj Quantum Inf* **4** 11. https://doi.org/10.1038/s41534-017-0055-x
[50] Saggio V, Dimic A, Greganti C, Rozema L A, Walther P and Dakic B 2019 Experimental few-copy multipartite entanglement detection *Nat. Phys.* **15** 935-40. https://doi.org/10.1038/s41567-019-0550-4
[51] Zhang R Q, Hou Z B, Tang J F, Shang J W, Zhu H J, Xiang G Y, Li C F and Guo G C 2022 Efficient experimental verification of quantum gates with local operations *Phys. Rev. Lett.* **128** 020502. https://doi.org/10.1103/PhysRevLett.128.020502
[52] Yu X-D, Shang J and Gühne O 2021 Statistical methods for quantum state verification and fidelity estimation *arXiv e-prints* arXiv:2109.10805. https://doi.org/arXiv:2109.10805
[53] Tison C C, Steidle J A, Fanto M L, Wang Z, Mogent N A, Rizzo A, Preble S F and Alsing P M 2017 Path to increasing the coincidence efficiency of integrated resonant photon sources *Opt. Express* **25** 33088-96. https://doi.org/10.1364/OE.25.033088
[54] Liu Y W, *et al.* 2020 High-spectra purity photon generation from a dual-interferometer-coupled silicon microring *Opt. Lett.* **45** 73-6. https://doi.org/10.1364/OL.45.000073

[55] Lu L L, *et al.* 2020 Three-dimensional entanglement on a silicon chip *npj Quantum Inf* **6** 30. https://doi.org/10.1038/s41534-020-0260-x